\documentclass[pra,twocolumn]{revtex4}

\usepackage{appendix}
\usepackage{graphicx}
\usepackage{amssymb,amsmath} 
\usepackage{epsfig}
\usepackage[utf8]{inputenc}
\usepackage[T1]{fontenc}
\usepackage{color}
\usepackage{soul}

\def\B#1{\left(#1\right)}
\def\BB#1{\left[#1\right]}

\def\be{\begin{equation}}
\def\ee{\end{equation}}

\def\bee{\begin{equation*}}
\def\eee{\end{equation*}}

\begin{document}

\title{Numerical studies of ground state fidelity of the Bose-Hubbard model}
\author{Mateusz \L\k{a}cki$^1$, Bogdan Damski$^1$, and Jakub Zakrzewski$^{1,2}$} 
\affiliation{\mbox{$^1$Institute of Physics, Jagiellonian University, Reymonta 4, 30-059 Krak\'ow, Poland}
\mbox{$^2$Mark Kac Complex Systems Research Center, Jagiellonian University, Reymonta 4, 30-059 Krak\'ow, Poland}
}
\begin{abstract}

We compute ground state fidelity of the one-dimensional Bose-Hubbard model at unit
filling factor. 
To this aim, we apply the DMRG algorithm to systems with open and
periodic boundary conditions. 
We find that fidelity differs significantly in the two cases and
study its scaling properties  in the quantum critical regime. 
\end{abstract}
\maketitle

\section{Introduction}
\label{sec_introduction}

Ground state fidelity is defined as the overlap between two ground states \cite{Zanardi}
\be
F(\lambda,\delta) = \left|\langle \lambda-\delta/2|\lambda+\delta/2\rangle\right|,
\label{F}
\ee
where $|\lambda\rangle$ is the ground state of the Hamiltonian
$\hat H(\lambda)$, $\lambda$ is the parameter of that Hamiltonian, and $\delta$
is its shift. Since fidelity quantifies similarity of 
the ground states, it  is a useful probe of quantum criticality.

Indeed, the quantum phase transition happens when the ground state of the system 
can be fundamentally changed by  a small variation of some parameter of the
system's Hamiltonian \cite{Sachdev,ContinentinoBook,SachdevToday}. 
Since the ground states belonging to different phases
 have little in common, fidelity is expected to  exhibit a marked drop 
across the critical point \cite{Zanardi}. 
This intuitive remark was studied in a surprisingly-large variety of physical models,
for example, in numerous spin models (Ising, XY, XYX, XXZ, Heisenberg, Kitaev, Lipkin-Glick-Meshkov, etc.)  \cite{GuReview}.

The scaling theory of quantum phase transitions was employed to 
predict the dependence of fidelity on the distance $|\lambda-\lambda_c|$ from the critical 
point, the system size $M$, and the critical exponent $\nu$ characterizing 
the power-law divergence of the correlation length $\xi\sim|\lambda-\lambda_c|^{-\nu}$
\cite{ABQ2010,Polkovnikov,BDfid1}. 
The key insights provided by these studies  can be briefly summarized 
as follows.

In the limit of $\delta\to0$ taken under the fixed system size $M$ fidelity 
can be expanded as 
\be
F(\lambda,\delta)=1-\chi(\lambda)\frac{\delta^2}{2} + O\B{\delta^4}.
\label{fidsus}
\ee
Then, one finds that around the critical point $\chi \sim M^{2/d\nu}$, while far 
away from it $\chi\sim M/|\lambda-\lambda_c|^{2-d\nu}$ \cite{ABQ2010,Polkovnikov}.
$\chi$ is known as fidelity susceptibility and  $d$ stands for the system's
dimensionality.

On the other hand, in the limit of $M\to\infty$ taken at the fixed field shift
$\delta$ one finds that near the critical point $\ln F \sim -M|\delta|^{d\nu}$,
while far away from it $\ln F \sim -M\delta^2/|\lambda-\lambda_c|^{2-d\nu}$ 
\cite{BDfid1}; see also Ref.~\cite{Zhou} for a
similar approach to fidelity in the  thermodynamically-large systems and Ref.
\cite{BDfid2} for some modifications to these scaling laws.

We will study below the Bose-Hubbard model at unit filling factor
\cite{BHearly,Fisher89}. 
This model describes  interacting bosons in a lattice. It can be experimentally realized in 
optical lattices filled with 
ultra cold atoms. This was proposed in Ref. \cite{JakschPRL1998} and accomplished in a cubic lattice a few years 
later \cite{Greiner2002}. Soon by the appropriate
modifications of the  lattice potential, a one-dimensional (1D) 
version of the model was also realized \cite{Stoferle2004}. Since then the Bose-Hubbard model
and its generalizations form  standard starting points in describing cold atoms
in periodic potentials (see Ref. \cite{Lewenstein12} for a recent review). 

The Bose-Hubbard model allows for  
the quantum phase transition between Mott insulator and superfluid states.
Importantly, this transition is of  
Berezinskii-Kosterlitz-Thouless (BKT) type \cite{Fisher89}.
The correlation length is infinite on the superfluid side 
and it  exponentially diverges near the critical point  on the Mott insulator side:
$\ln\xi\sim1/\sqrt{\lambda_c-\lambda}$ \cite{PaiPRL1996,Monien1998}. 
This means that no critical exponent $\nu$ can  be defined on either side of the transition 
and so the above scaling expressions cannot be directly used.

As far as we know,
two papers report results on fidelity of the Bose-Hubbard model 
\cite{Buonsante2007,Rigol2013}. Both of them  discuss numerical simulations
showing that the minimum of fidelity is 
strongly shifted from the critical point even in systems composed of
a few hundreds of atoms (at unit filling factor). More importantly, no convincing argument
for the extrapolation of the position of the critical point from the location of the 
minimum of fidelity in finite-size systems has been proposed so far.  
Therefore, the understanding of fidelity of the Bose-Hubbard model is
incomplete, which  motivates our numerical ``experiment'' on this model. 
We consider systems larger than those previously studied, describe an 
unexpected sensitivity of fidelity to the boundary conditions,
and systematically study fidelity around its minimum.

\section{Model}
We study fidelity of the Bose-Hubbard model:
\bee
\hat H = - J\sum_{i=1}^M \B{\hat a_i^\dag\hat a_{i+1} + {\rm h.c.}}
+\frac{U}{2}\sum_{i=1}^M \hat n_i\B{\hat n_i-1},
\eee
where $M$ is the number of lattice sites, $\hat n_i=\hat a^\dag_i\hat a_i$,
and the creation/annihilation operators satisfy the bosonic commutation relations.
The first term describes 
the tunnelling between lattice sites, while the second one accounts for
on-site interactions \cite{homo}. The phase diagram of this model
depends on the filling factor $N/M$ and  the $J/U$ ratio 
($N$ is the number of atoms in a lattice). 
For non-integer filling factors the system is always superfluid.
When $N/M$ is integer, the system is in the Mott insulator
phase for $J/U < (J/U)_c$ and in the superfluid phase for $J/U > (J/U)_c$. 
The position of the critical point was studied in numerous theoretical papers (see
Sec. II of Ref.~\cite{Rigol2013} for the recent survey of these studies).
The reported values for $(J/U)_c$ range from about $0.27$ to about $0.3$.

The spread of these estimations clearly highlights the complexity of the 1D
Bose-Hubbard model, which  unlike its famous
fermionic cousin \cite{Essler}, is not integrable. Thus, its
analytical studies are necessarily approximate. We will discuss below some
relevant approximations and critically evaluate their 
applicability to the computation of fidelity.

The first simplification one can invoke is the Taylor  expansion (\ref{fidsus}),
where the $\lambda$ and $\delta$ dependence of fidelity separate out. One is then
left with the computation of fidelity susceptibility, which can be exactly written as \cite{GuReview}
\be
\chi(\lambda)= \sum_{S\neq0} 
\frac{|\langle \phi_S(\lambda)|\hat V|\phi_0(\lambda)\rangle|^2}{\BB{E_0(\lambda) - E_S(\lambda)}^2},
\label{perturb}
\ee
where $\hat V = \partial\hat H/\partial\lambda$ and $|\phi_S(\lambda)\rangle$ 
is an eigenstate of $\hat H(\lambda)$
to the eigenvalue $E_S(\lambda)$ ($S\equiv0$ corresponds to the ground state).
Eq.~(\ref{perturb}) can be in principle exactly evaluated when the
eigenstates $|\phi_S(\lambda)\rangle$ and their eigenenergies $E_S(\lambda)$ are exactly known.
This is possible in
the Bose-Hubbard model only for $\lambda=J/U=0$ (deep Mott insulator limit)
and $\lambda=U/J=0$ (deep superfluid limit).

\begin{figure}
\includegraphics[width=\columnwidth,clip=true]{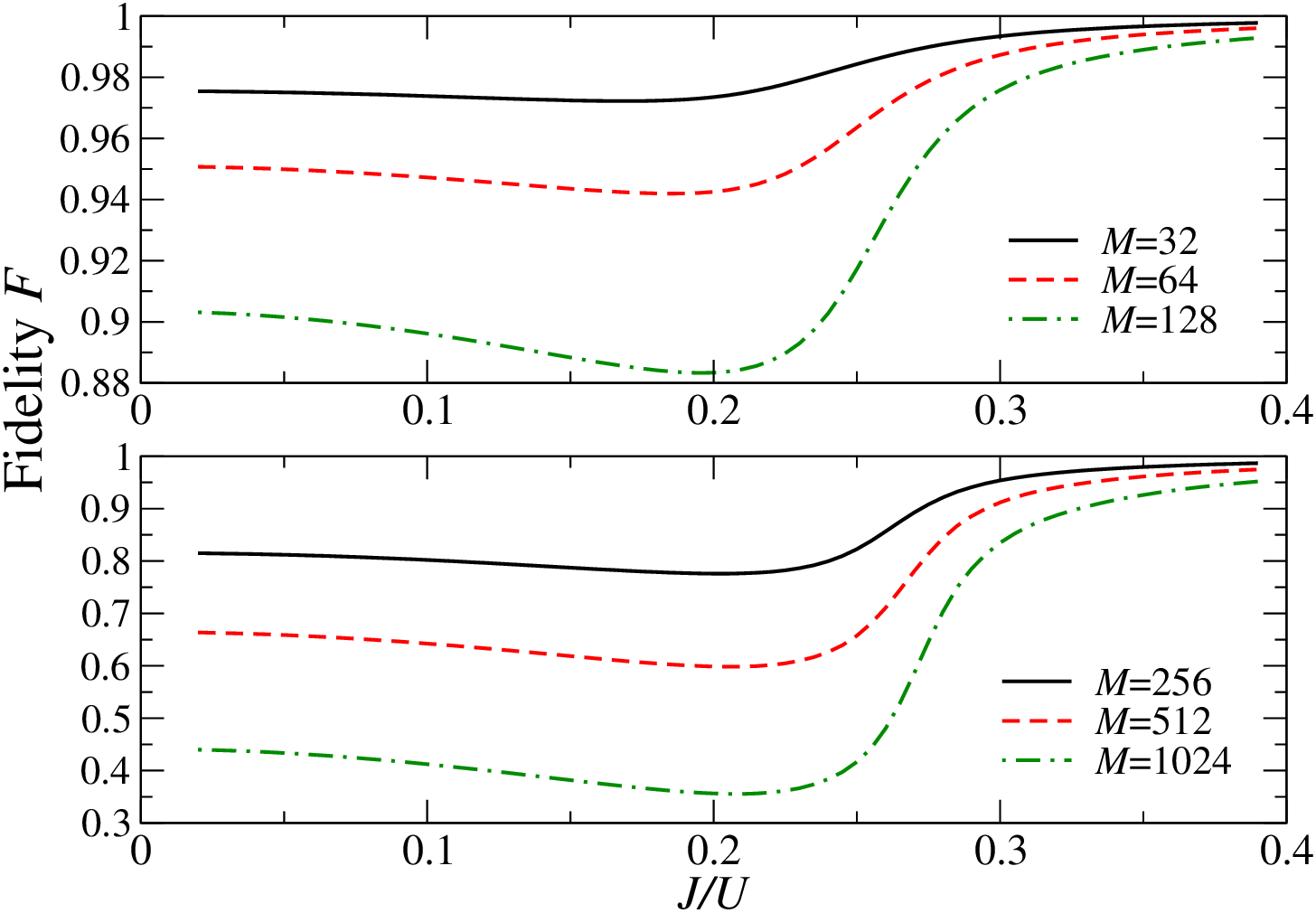}
\caption{(color online) Fidelity in an open Bose-Hubbard chain for different system sizes and the parameter shift $\delta=0.02$.
}
\label{open_delta0_02_Ndo1024}
\end{figure}

In the former case, one sets $\hat V= -\sum_{i=1}^M \hat a_i^\dag\hat a_{i+1} + {\rm h.c.}$ and
easily finds the eigenstates of $\hat H(J/U=0)=\frac{1}{2}\sum_{i=1}^M \hat n_i\B{\hat n_i-1}$, which 
immediately leads to 
\be
\chi(J/U=0)=2 \frac{N}{M}\B{\frac{N}{M}+1}M,
\label{chiJ0}
\ee
if the periodic boundary conditions and the integer filling factor are assumed.
Thus $\chi(J/U=0)$ is extensive, i.e., it scales linearly with the system size at the
fixed $N/M$.

In the latter case, a non-trivial result is obtained. We refer the reader to 
Appendix \ref{AppA} for the details of its derivation and quote here only the final expression:
\be
\chi(U/J=0) = \frac{N(N-1)}{5760M^2}\B{M^4+10M^2-11},
\label{chiU0}
\ee
valid for periodic boundary conditions and arbitrary filling factor $N/M$.
This result is super-extensive. The scaling $\chi(U/J=0)\sim M^4$ 
appears because  the low-energy spectrum of a non-interacting
bosonic gas in a lattice is quadratic in quasimomentum.
Noting that the
crossover from the quadratic to linear spectrum happens at \cite{Stoof2001}
\bee
\frac{U}{J}\sim \sin^2(k/2),
\eee
the qualitative departures from Eq.~(\ref{chiU0}) are
expected to happen at $U/J\sim \pi^2/M^2$.

For $J/U\neq0,\infty$, Eq. (\ref{perturb}) cannot be efficiently used to compute
fidelity, and  so its usefulness for  the analytical 
characterization of fidelity of the Bose-Hubbard 
model is very   limited.

\begin{figure}
\includegraphics[width=\columnwidth,clip=true]{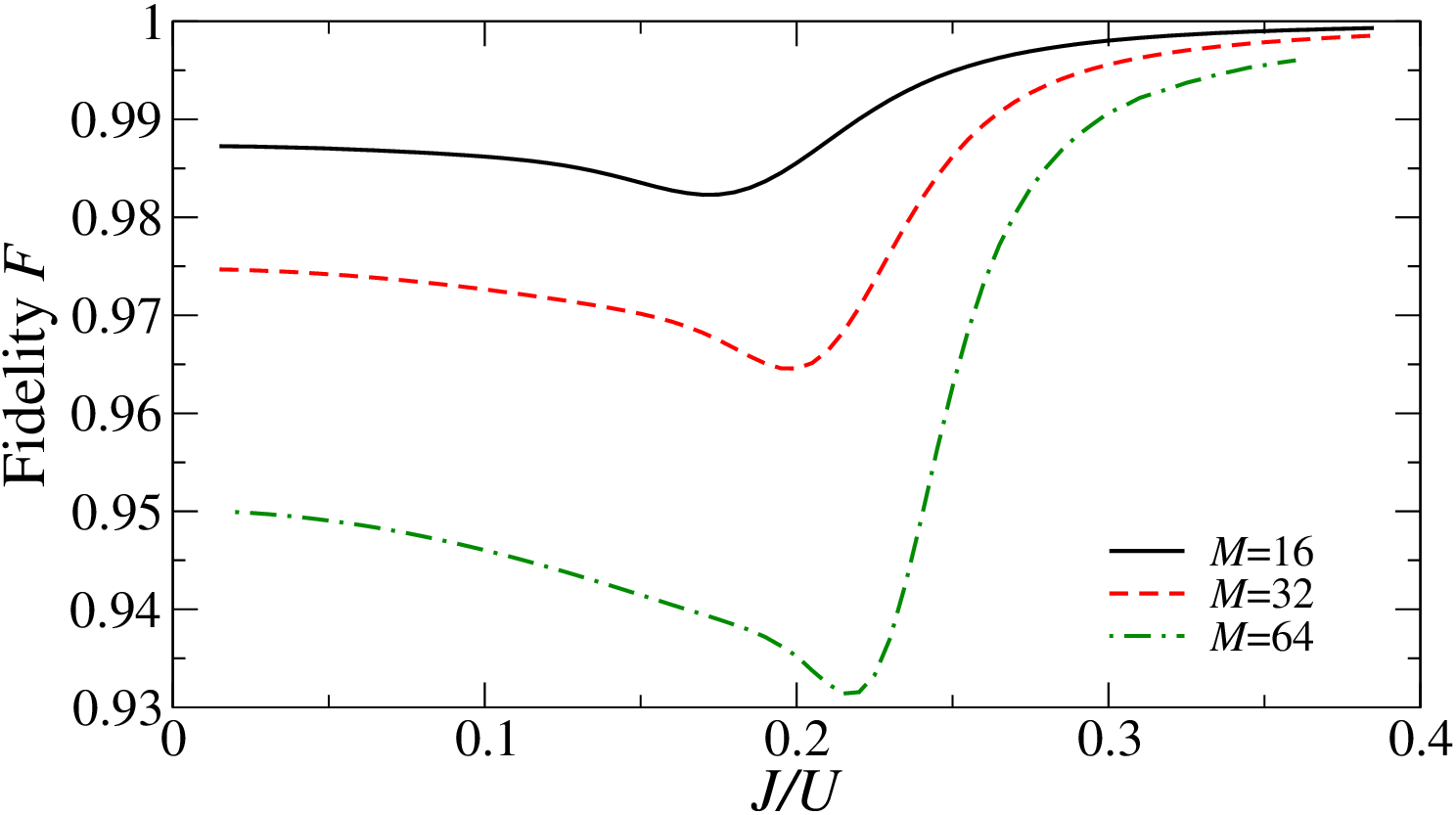}
\caption{(color online) 
Fidelity in a periodic Bose-Hubbard model for different system sizes and  $\delta=0.02$.
}
\label{per_delta0_02_Ndo64}
\end{figure}

The next approach in line that seems to be simple enough to provide some analytical 
insights about fidelity of the Bose-Hubbard model is the Bogolubov approach
\cite{Stoof2001}.
There are, however,  two problems associated with it. First, it does not 
describe  physics of the Mott insulator phase possessing a finite
excitation gap. Second, its validity even in the superfluid phase 
is questionable in 1D lattices due to the sizable population of the 
non-condensed atoms \cite{Stoof2001}. Since we are interested in the BKT transition of the
Bose-Hubbard model, our studies are restricted to the 1D model.

Finally, the  more advanced treatment of the superfluid phase is provided by the
Luttinger liquid theory; see e.g. Ref.~\cite{MonienLL} for the brief discussion of this  theory 
in the Bose-Hubbard context. Luttinger liquid theory can be regarded as an effective low energy
theory of systems, whose spectrum is linear in momentum and  whose correlations 
decay algebraically \cite{1D,Cazalilla2004}.
The computation of fidelity in the Luttinger liquid theory was 
presented in Refs.~\cite{LLfidel,LLfidel1,LLfidel2}. While Refs.
\cite{LLfidel,LLfidel1} provide a definite expression for fidelity,
Ref.~\cite{LLfidel2} argues that 
the zero temperature result depends on an arbitrary cut-off. This prediction
is verified by comparing the Luttinger liquid theory prediction for fidelity 
susceptibility to the actual result in the XXZ spin chain \cite{LLfidel2}.
Similarly as the Bose-Hubbard model, the XXZ model also undergoes a BKT transition. 
Therefore, we assume  that the findings of Ref. \cite{LLfidel2} are relevant for the
Bose-Hubbard model as well. 

Given all these complications, we focus on the 
numerics. We use Matrix Product State (MPS) techniques for both periodic 
($\hat a_{M+1}\equiv\hat a_{1}$) and open ($\hat a_{M+1}\equiv0$) 
boundary conditions. 
Boundary conditions significantly affect the complexity
of the numerical computations.

For open boundary conditions the usual DMRG algorithm can be used \cite{White1993,White1992,DMRG}.
This algorithm can be naturally formulated  in the MPS language  \cite{Schollwock2011}. 
It has allowed us to find  
the ground states of the Bose-Hubbard model for lattices containing up to 2048 
sites at the unit filling factor.
We have limited the bond dimension of the tensors forming the MPS 
representation to the 200 largest singular values . This has allowed us to 
obtain converged results for fidelity.

\begin{figure}
\includegraphics[width=\columnwidth,clip=true]{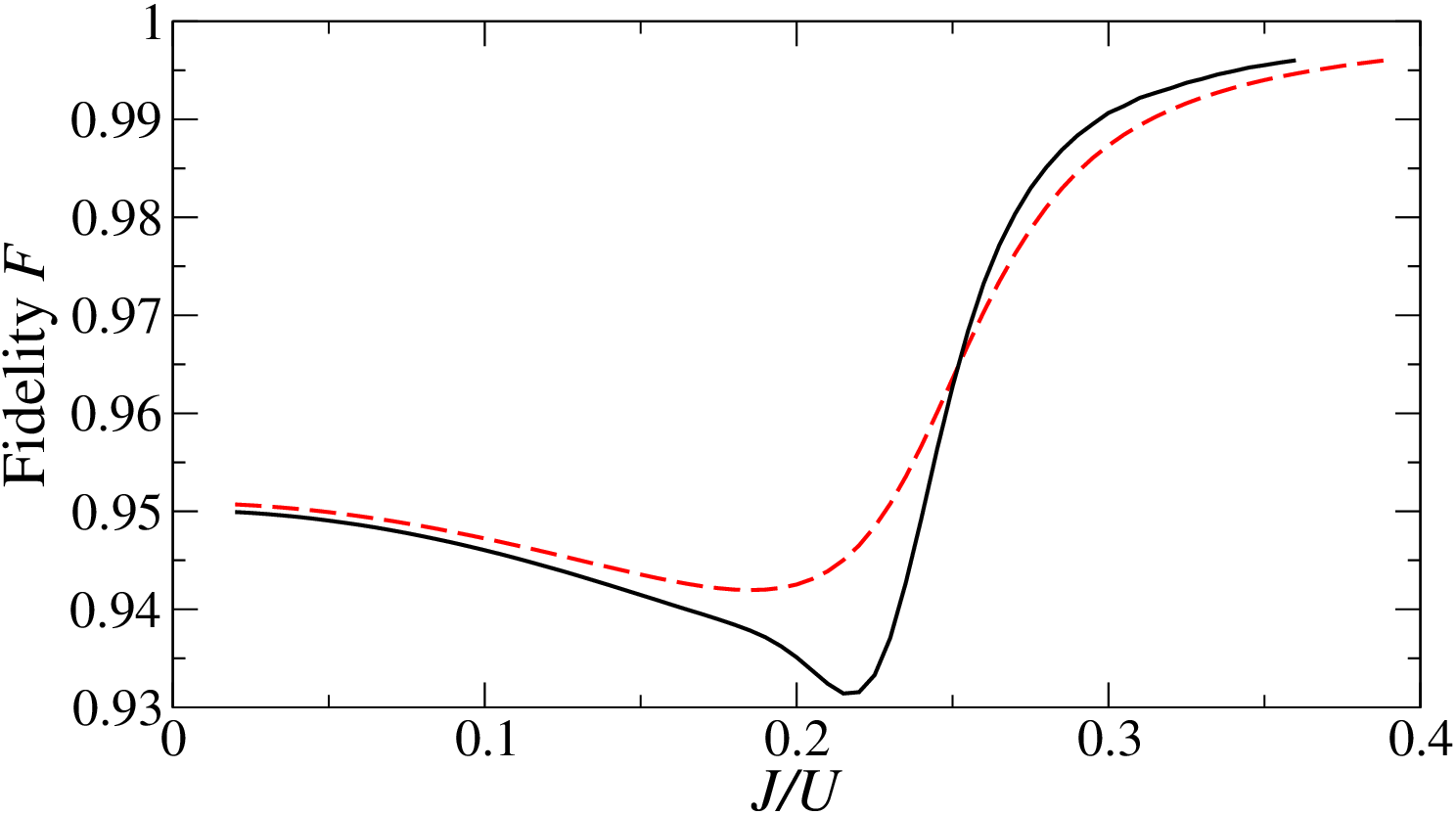}
\caption{(color online) Black solid line shows fidelity in a periodic chain,
while the dashed red line shows fidelity in an open chain. In both cases
$M=64$ and $\delta=0.02$.
}
\label{per_vs_open}
\end{figure}

To find the ground state for the periodic boundary conditions, we have used
imaginary time evolution technique \cite{KubaDom}. The
time evolution has been 
performed with the TEBD algorithm  \cite{Wall2009}.
In  periodic chains the singular values decrease more slowly than in the 
open ones \cite{Verstraete2004,Pippan2010}.
To obtain the converged fidelity, the bond dimension 220 of the MPS vectors 
was necessary even for relatively small periodic  systems consisting of 64 lattice sites with unit
filling. In both cases the local Hilbert space has been 
cut to the subspace allowing at most 6 particles per site. The accuracy of our
computations is discussed in Appendix \ref{AppB}.

Finally, we mention that for $M=12$ and both open and periodic chains 
we have computed fidelity via exact diagonalization and compared  such results
to the  DMRG numerics. The two approaches agree with each other.

Previous  studies of fidelity of the Bose-Hubbard model at unit filling factor
were restricted to $M\le12$ on 
periodic lattices (exact diagonalization) \cite{Buonsante2007} and $M\le120$ 
on the open lattices (DMRG) \cite{Rigol2013}. Ref.~\cite{Buonsante2007} reports
results on both fidelity and fidelity susceptibility, while Ref.
\cite{Rigol2013} focuses on fidelity susceptibility.

\section{Fidelity}
\label{sec_fidelity}
From now on, we set $N=M$, i.e., we study the unit filling case.
Typical results that we obtain in open and periodic systems 
are presented in Figs. \ref{open_delta0_02_Ndo1024} and
\ref{per_delta0_02_Ndo64}, respectively.

First, we notice that the minimum of fidelity lies on the Mott 
insulator side. This is an expected feature because 
the ground states near the critical point change more rapidly on
the Mott insulator side of the transition (see e.g. the 
exponential dependence of the correlation
length in the Mott phase near the critical point).

\begin{figure}
\includegraphics[width=\columnwidth,clip=true]{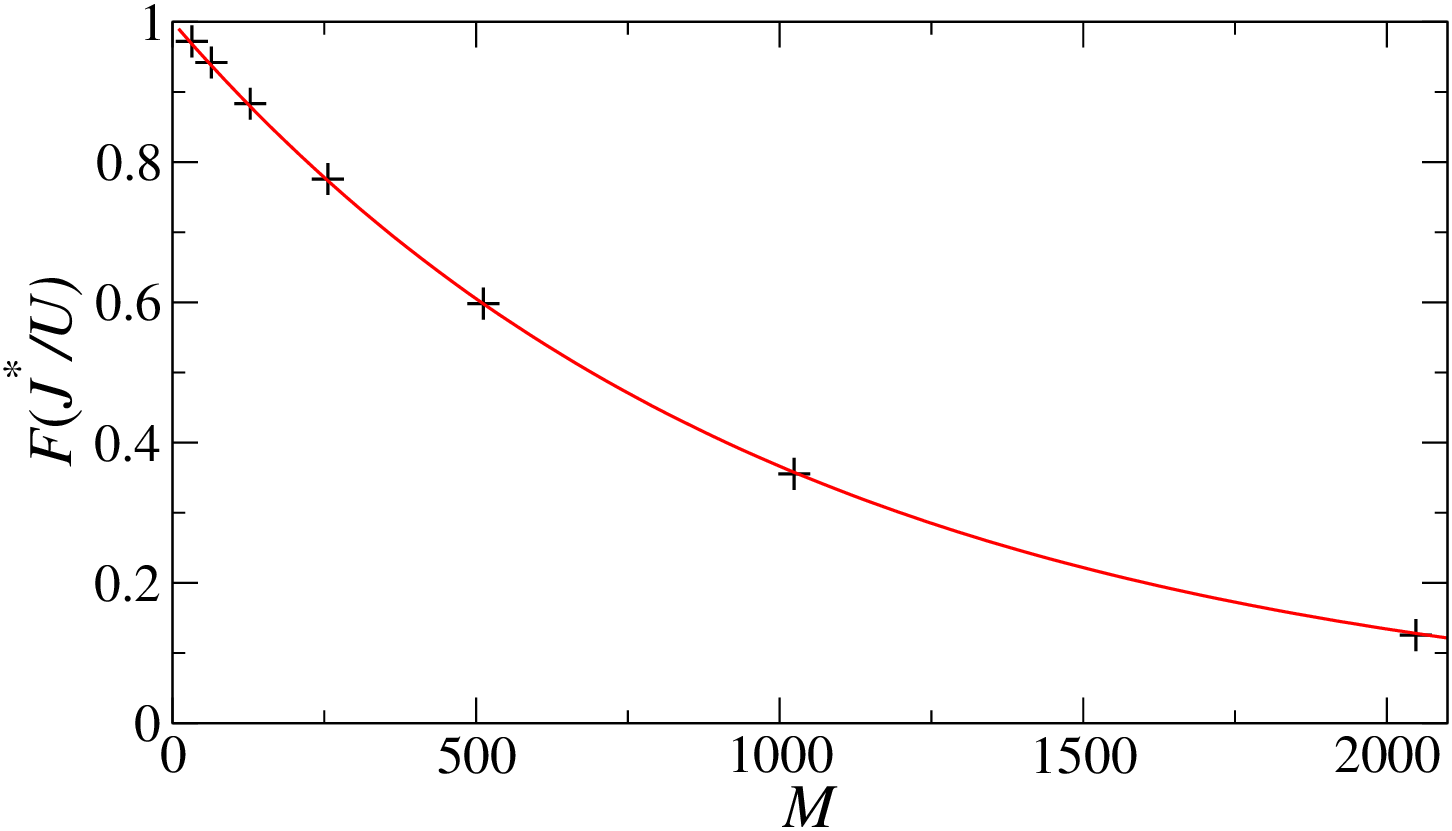}
\caption{(color online) The minimum of fidelity as a function of the system size 
(open boundary conditions). Pluses come from numerics done for $\delta=0.02$
and $M=32,64,128,256,512,1024,2048$. 
The line represents Eq.~(\ref{FJstar}) with $\alpha\simeq2.51$ obtained from the fit.
}
\label{Fwmin_0_02_open}
\end{figure}

Second, there are easy-to-notice differences between the periodic and open
chain results. As shown in Fig. \ref{per_vs_open}, the minimum in the open chain 
is much more shallow and more distant from the critical point than the one in
the  periodic chain. As expected, periodic chains probe quantum criticality more robustly. 
However, the magnitude of the difference between the two cases  is surprising.
It  prompts separate numerical studies of the periodic and open chains.

\section{Open boundary conditions}
\label{sec_open}
We start the discussion from looking at the value of fidelity 
at the minimum. We denote the position of the minimum as $J^*/U$.
Our numerics supports the following  expression  
\be
F(J^*/U,\delta) = \exp\B{-\alpha M\delta^2},
\label{FJstar}
\ee
where $\alpha$ is some constant. Note that since we work with $N/M=1$, it is
impossible to decide whether there should be $M$ or $N$ in Eq. (\ref{FJstar}).

\begin{figure}
\includegraphics[width=\columnwidth,clip=true]{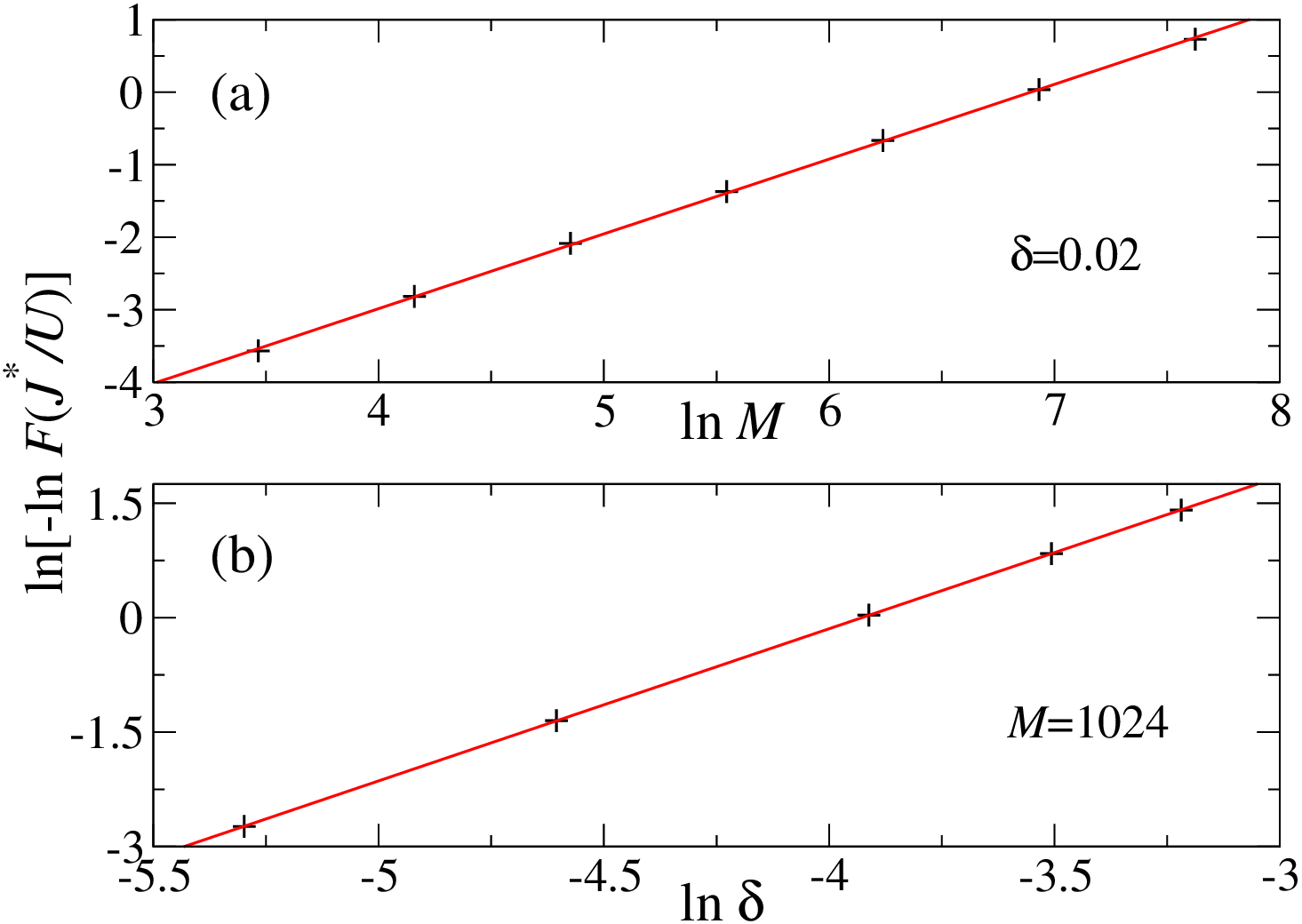}
\caption{(color online) Scaling of the minimum of fidelity with the 
system size $M$ and the parameter shift $\delta$ (open boundary conditions). 
Panel (a):  Pluses come from numerics done for 
$M=32,64,128,256,512,1024,2048$ \cite{remark_spline}. 
The line represents the  fit:  $\ln\BB{-\ln F\B{J^*/U}} = -7.142 + 1.032 \ln M$. 
Panel (b): Pluses come from numerics done for 
$\delta=0.005, 0.01, 0.02, 0.03, 0.04$ \cite{remark_spline}, 
while the line represents the fit:
$\ln\BB{-\ln F\B{J^*/U}} = 7.833  + 1.994 \ln\delta$. 
Both fits  accurately support the  key features of Eq.~(\ref{FJstar}): 
$\ln F\B{J^*/U} \sim -M$ tested in panel (a) and $\ln F\B{J^*/U}\sim-\delta^2$
tested in panel (b).
}
\label{Fwmin_Ln_open}
\end{figure}

The numerical evidence supporting Eq.  (\ref{FJstar}) is discussed in detail  in Figs.
\ref{Fwmin_0_02_open} and \ref{Fwmin_Ln_open}. It is worth to stress that
Eq.  (\ref{FJstar}) works in these figures also when $F(J^*/U,\delta)\ll1$, i.e., 
when fidelity susceptibility fails to account for fidelity.

The natural question arising now is the following: Can we extrapolate the position of the
critical point by studying the scaling of $J^*/U$ with the system size $M$? 

We were unable to find an extrapolation scheme that would give us the correct
location of the critical point.  We will mention below two attempts.

\begin{figure}
\includegraphics[width=\columnwidth,clip=true]{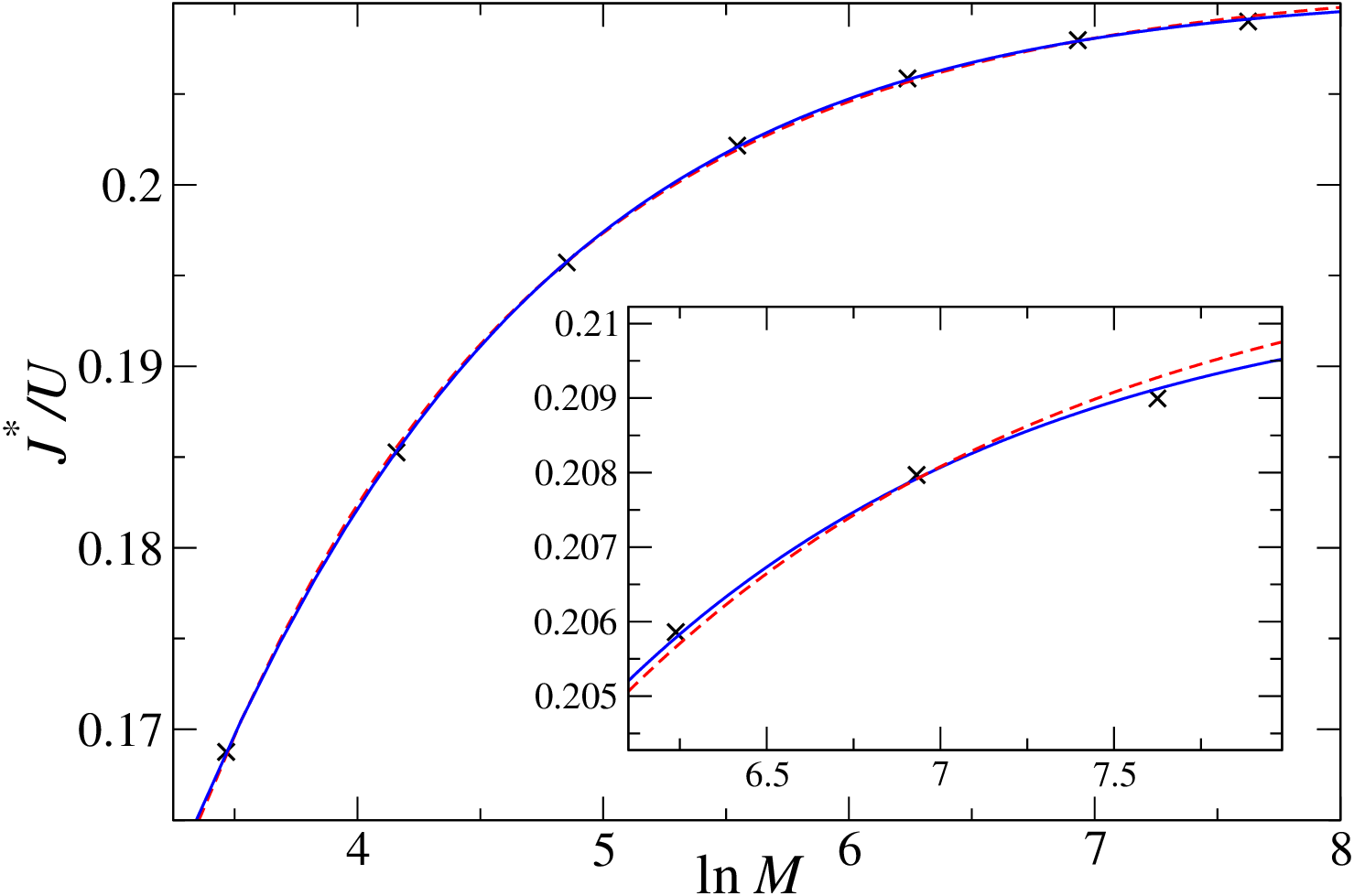}
\caption{(color online) Position of the minimum of fidelity as a function of
the system size (open boundary conditions). 
X's show numerics done for $M=32,64,128,256,512,1024,2048$ and 
$\delta=0.02$ \cite{remark_spline}. The red (dashed) line is the fit
(\ref{Jfit}) while the blue (solid) corresponds to (\ref{Jfit_ln}). The inset shows the zoom for large $M$ to enable comparison of both fits.
}
\label{Jstart_delta0_02_open}
\end{figure}

First, we have tested 
\be
\frac{J^*(M)}{U} = \frac{J^*(\infty)}{U} - \frac{a}{M^b},
\label{Jfit}
\ee
with fitted parameters 
$J^*(\infty)/U=0.2114\pm0.0003$, $a=0.53\pm0.02$, and $b=0.73\pm0.02$ 
obtained via standard {\it Mathematica} fitting procedure
(Fig. \ref{Jstart_delta0_02_open}). Thus the extrapolated position of
the minimum in the infinite system, $J^*(\infty)/U\simeq0.2114$, 
is nowhere near the expected location of the critical point  $(J/U)_c$. 
The fit (\ref{Jfit}) was proposed in Ref. \cite{Buonsante2007} studying
fidelity of the Bose-Hubbard model, 
but no theory supporting it was discussed there. 
This  fit was also  used  to extrapolate the position  of the maximum of fidelity susceptibility in the 2D 
Ising model in  a transverse field \cite{ABQ2010}.  The fit provided  the correct 
location of the critical point and the critical exponent $\nu$
(following Ref. \cite{Paris}, it was verified in Ref. \cite{ABQ2010} that $b=1/\nu$ in the
2D Ising model in a transverse field; in a 1D Ising model in a transverse field
$b=2/\nu$, which was shown in Refs. \cite{Paris,BDfid3}).  In the Bose-Hubbard
model, however, the critical exponent $\nu$ is undefined and it is unclear to
us how to improve the fit (\ref{Jfit}) to properly extrapolate the position of
the critical point.

Second, we tried 
\be
\frac{J^*(M)}{U} = \frac{J^*(\infty)}{U} - a\frac{\ln M}{M^b}
\label{Jfit_ln}
\ee
to include logarithmic finite-size corrections near the BKT
transition (Fig. \ref{Jstart_delta0_02_open}). 
 The fit has provided  $J^*(\infty)/U=0.2106\pm0.0001$, $a=0.375\pm0.007$, and $b=0.991\pm0.006$. 
While  this value for $J^*(\infty)/U$  is very close to that obtained previously,  
the quality of this fit is  better (the inset of Fig.
\ref{Jstart_delta0_02_open}).  This suggests that the 
finite-size correction to the position of the minimum of fidelity scales as 
$\ln M/M$. To quantify the difference between the fits, we provide 
chi-squared, i.e., the sum of the squared differences between the fitted curve and
the numerical data. It equals about $2.7\times10^{-7}$ for the fit (\ref{Jfit})
and $3.9\times10^{-8}$ for the fit (\ref{Jfit_ln}). Thus, the fit (\ref{Jfit_ln}) is 
indeed  better than the (\ref{Jfit}) one.

\begin{figure}
\includegraphics[width=\columnwidth,clip=true]{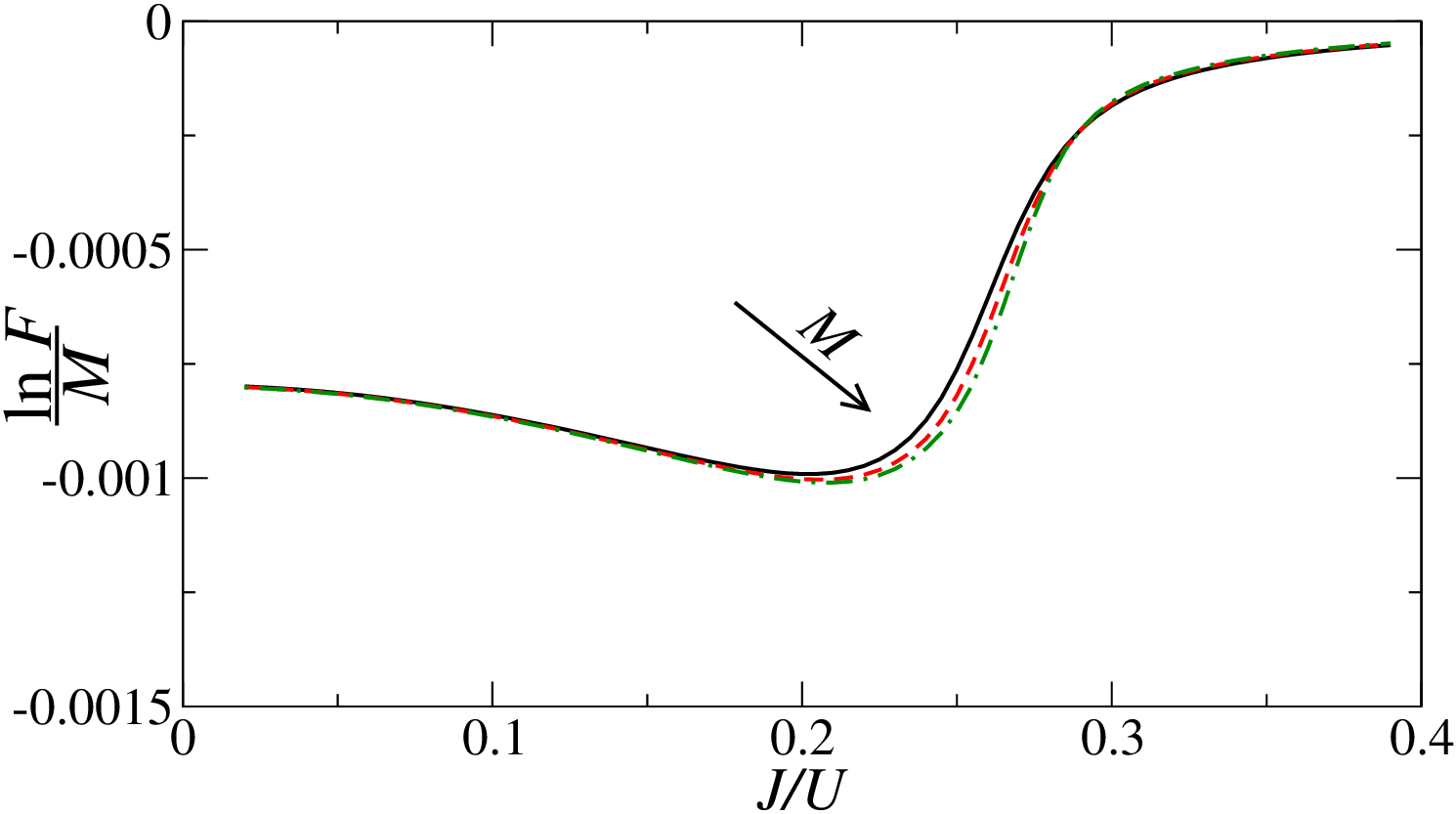}
\caption{(color online) Fidelity per site for $\delta=0.02$ 
and three system sizes: $M=256,512,1024$ (black solid, red dashed, green dot-dashed, respectively). 
This plot is made from data from the lower panel of Fig. \ref{open_delta0_02_Ndo1024}.
}
\label{lnF_M}
\end{figure}

Finally, we look closer at fidelity per site, i.e., $\ln F/M$, which is plotted in Fig. \ref{lnF_M}.
As discussed in Sec. \ref{sec_introduction}, this quantity is expected to have
a finite non-zero value in the thermodynamic limit in the systems, where the
correlation length diverges algebraically. This was predicted and
observed  in several models \cite{Zhou,BDfid1,SenPRB2012,Adamski2013}. It is unclear from Fig. \ref{lnF_M}
whether the  same holds for the Bose-Hubbard model. Further studies
focusing on larger systems are  needed to settle the system-size 
dependence of fidelity per site.

Next, we turn our attention to the fidelity susceptibility $\chi(J/U)$.
Since fidelity (\ref{F}) is symmetric with
respect to the $\delta\to-\delta$ transformation, we compute 
\bee
2\frac{1-F(J/U,\delta)}{\delta^2}
\eee
for  $\delta=0.005,0.01$ and all $J/U$'s of interest and fit it  with 
\be
\chi(J/U) + \psi(J/U)\delta^2,
\label{extrap}
\ee
where $\chi(J/U)$ and $\psi(J/U)$ are the fitting parameters.
We plot such obtained fidelity susceptibility in Fig.
\ref{fidsus_extrap_open}.

The deep Mott insulator regime, $J/U\to0$, can be computed exactly.
For open boundary conditions we get 
$\chi(0)=4(M-1)$ at the unit filling factor.

Similarly as in Ref.~\cite{Rigol2013}, we observe that
the $\chi/M$ curves cross near the critical point, i.e., around $J/U\simeq0.287$ 
(the inset of Fig. \ref{fidsus_extrap_open}). Indeed, 
we estimate  from Fig. 2 of Ref.~\cite{Rigol2013} that the crossing occurs there at
 $J/U\simeq 1/3.5\simeq0.286$, which is completely consistent with  our result.
One could speculate that the position of the critical point might be linked to
such  a crossing, i.e., to the point where $\chi\sim M$. One should be careful,
however, as such a crossing is absent in periodic systems
discussed in Sec. \ref{sec_periodic}.

We have also  studied the position of the maximum of fidelity
susceptibility. For example, setting $\delta=0.02$ as in Fig. \ref{Jstart_delta0_02_open},
we have found that the minima of fidelity and maxima of fidelity
susceptibility roughly coincide (as expected, the larger the system size is,
the bigger the discrepancy is). We have fitted the position of the maximum 
of fidelity susceptibility for $M=32,64,128,256,512,1024,2048$ 
with the power-law (\ref{Jfit}) and again obtained
$J^*(\infty)/U =0.2121\pm0.0002$.

\section{Periodic boundary conditions}
\label{sec_periodic}
This section presents our results on fidelity in the periodic Bose-Hubbard model.
Due to the numerical limitations, the system sizes studied  here are
restricted to $M\le64$, which is the factor of $2^5$ smaller 
than the range of the system sizes considered in Sec. \ref{sec_open}, but the
factor of $2^2$ larger than the system sizes considered  in Ref. \cite{Buonsante2007}.

\begin{figure}
\includegraphics[width=\columnwidth,clip=true]{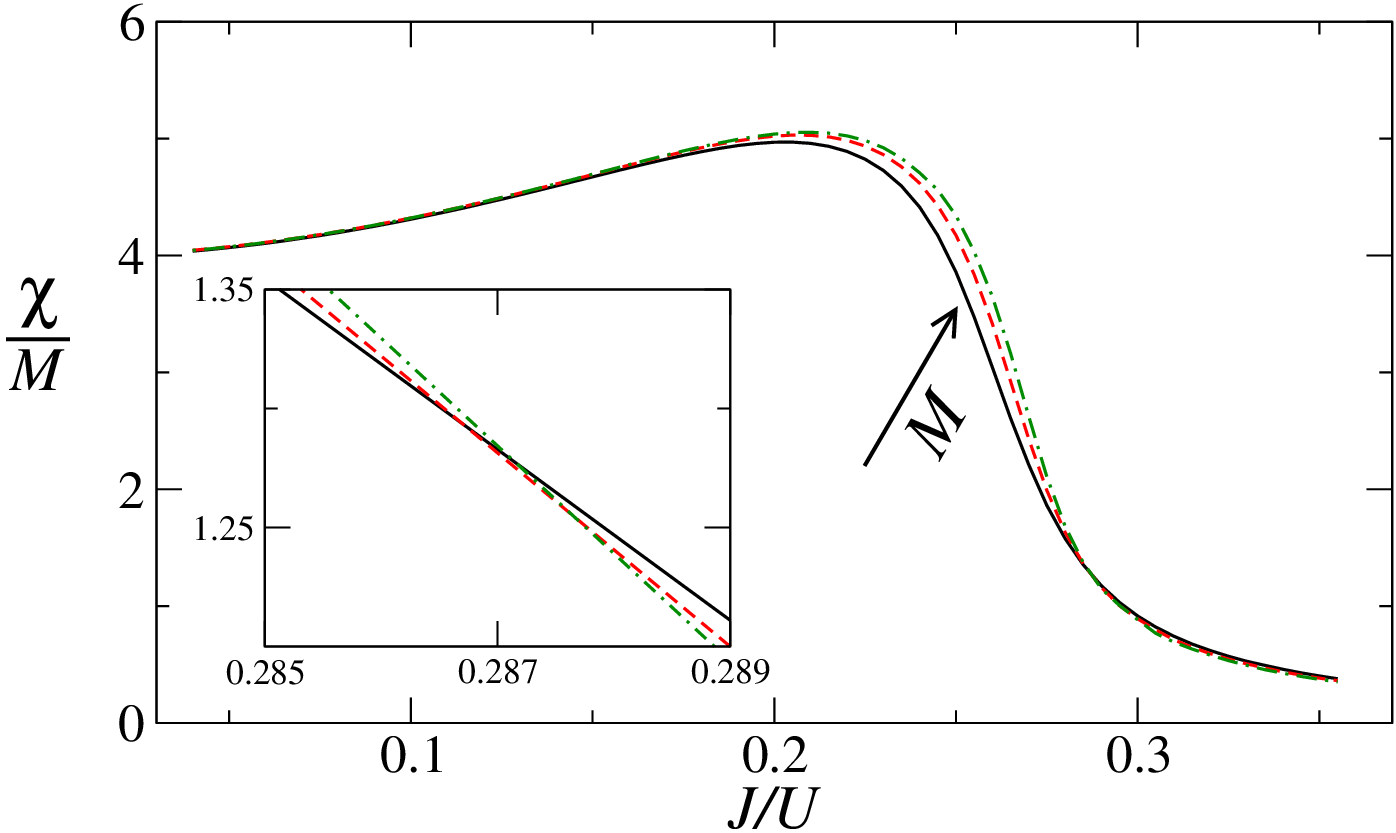}
\caption{(color online) Fidelity susceptibility per site (open boundary conditions). 
The black solid, red dashed, green dot-dashed lines represent data for 
$M=256,512, 1024$, respectively.
The inset enlarges the part of the main plot, where the curves cross.
}
\label{fidsus_extrap_open}
\end{figure}

\begin{figure}
\includegraphics[width=\columnwidth,clip=true]{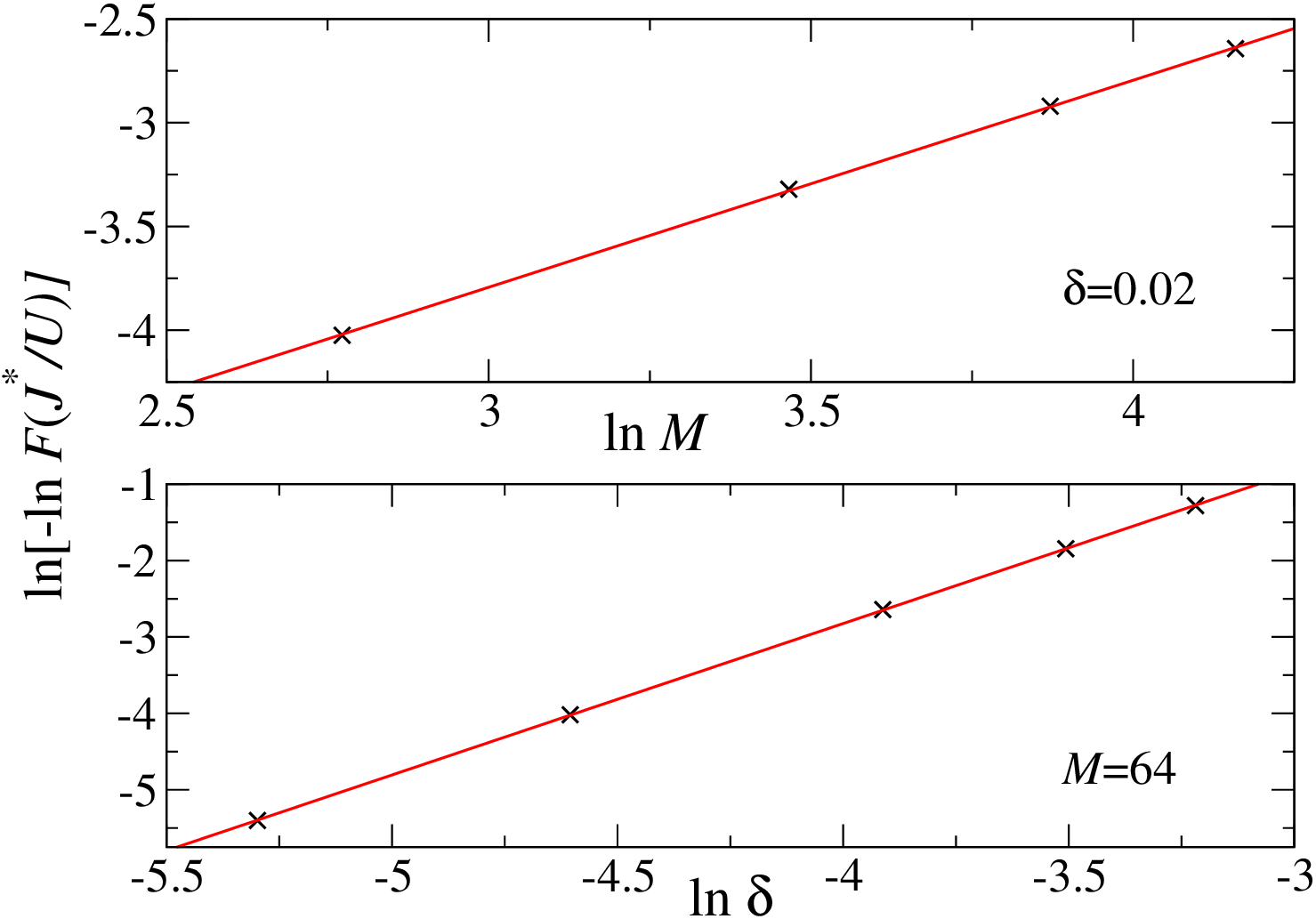}
\caption{(color online) Scaling of the minimum of fidelity with the system size $M$
and the parameter shift $\delta$ in a periodic Bose-Hubbard chain. 
Upper panel: X's show numerics for $M=16,32,48,64$ \cite{remark_spline}. The line represents the fit 
$\ln\BB{-\ln F\B{J^*/U}} = -6.787 + 0.998\ln M$. 
Lower panel: X's show numerics for $\delta =0.005,0.01,0.02,0.03,0.04$.
The line represents the fit  $\ln\BB{-\ln F\B{J^*/U}} = 5.104 +  1.982\ln\delta$.
}
\label{periodic_M64_spline}
\end{figure}

As in Sec. \ref{sec_open}, we find numerical evidence towards the
proposed scaling of the position of the minimum of fidelity (\ref{FJstar}). This is presented 
in detail in Fig. \ref{periodic_M64_spline}. There is no significant difference here
between the open and periodic results, compare Figs. \ref{Fwmin_Ln_open} and 
\ref{periodic_M64_spline}.

\begin{figure}
\includegraphics[width=\columnwidth,clip=true]{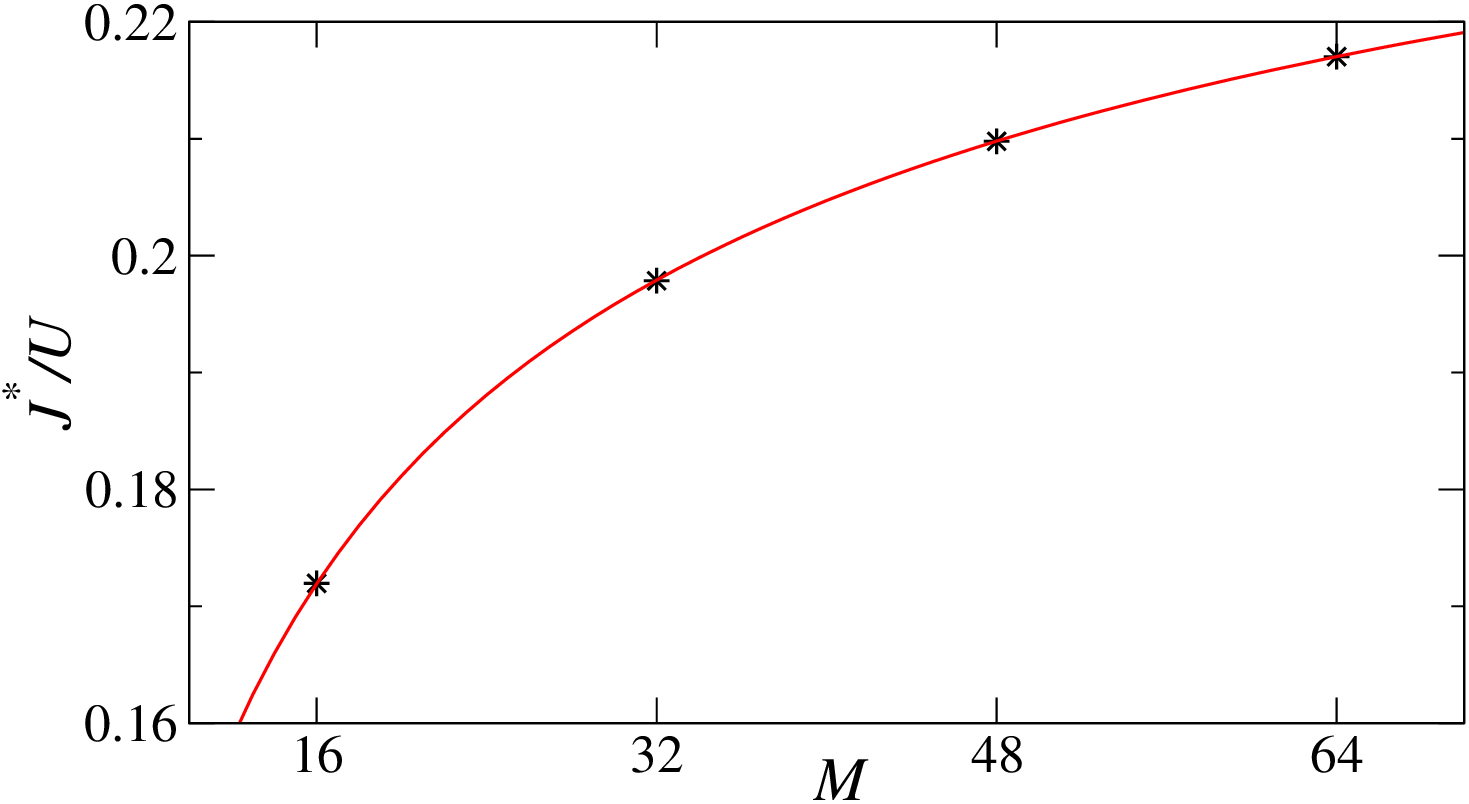}
\caption{(color online) Position of the minimum of fidelity 
as a function of
the system size $M$ in the periodic Bose-Hubbard model. Stars show numerics done
for $\delta=0.02$ \cite{remark_spline}, 
while the line represents Eq.~(\ref{Jfit}) with 
$J^*(\infty)/U\simeq0.27$, $a\simeq0.33$, and $b\simeq0.44$ (all coming from
the fit).
}
\label{Jstar_delta0_02_periodic}
\end{figure}

On the other hand, the extrapolation of the position of the minimum of
fidelity through  Eq.~(\ref{Jfit}) provides a markedly different result with
respect to what we have found in Sec. \ref{sec_open}. This extrapolation is 
discussed in Fig. \ref{Jstar_delta0_02_periodic}. The extrapolated 
 position of the minimum in an infinite system is $J^*(\infty)/U=0.270\pm0.008$, 
which agrees with some of the previous estimations of the position of the critical 
point \cite{Rigol2013}. We see also from the fit 
that the convergence with the system size to the asymptotic result is 
slow: $b=0.44\pm0.05$ in formula (\ref{Jfit}).


\begin{figure}
\includegraphics[width=\columnwidth,clip=true]{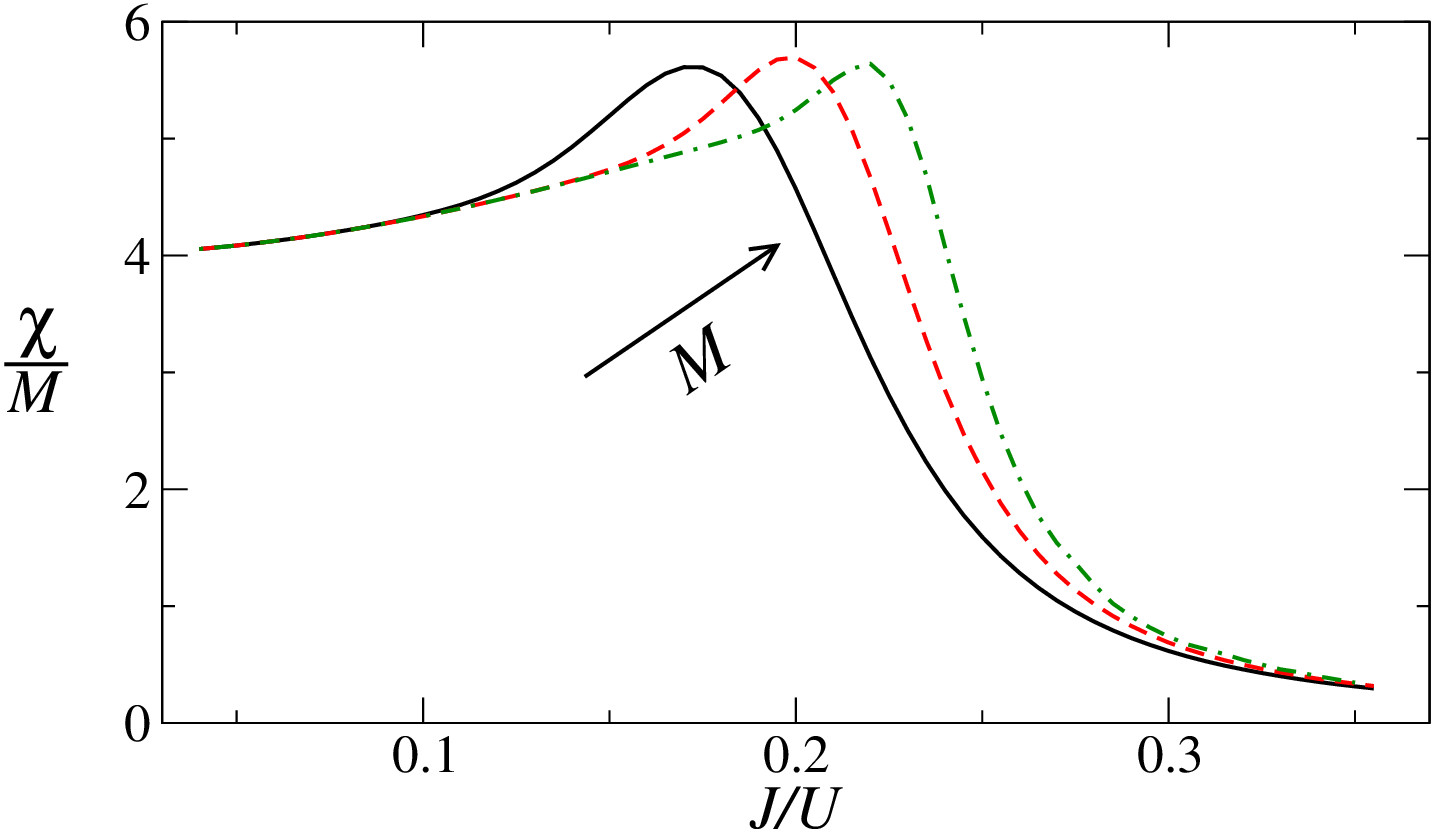}
\caption{(color online) Fidelity susceptibility per site (periodic boundary conditions). 
The black solid, red dashed, green dot-dashed lines represent data
for $M=16,32, 64$, respectively. 
Fidelity susceptibility is obtained through the extrapolation procedure discussed around Eq. (\ref{extrap}). 
}
\label{fidsus_extrap_periodic}
\end{figure}

 Next, we focus on fidelity susceptibility $\chi(J/U)$,
which is plotted   in Fig. \ref{fidsus_extrap_periodic}.
First, we notice again that $\chi/M\to4$ as $J/U\to0$. This follows from Eq.
(\ref{chiJ0}) taken at $N/M=1$. Moreover, it is also seen that $\chi/M$ is about the same 
at the maximum for all three curves ($M=16,32,64$).
Second, we see from Fig. \ref{fidsus_extrap_periodic} 
that the fidelity susceptibility curves
do not cross near the critical point in the periodic chain. However, they do cross near the critical point in 
the open chain (the inset of Fig. \ref{fidsus_extrap_open}). This  suggests that
the inhomogeneities appearing near the edges in the open chain are 
responsible for the crossing in the open system
(see also  Sec. \ref{sec_discussion}). If this is indeed the case, it is even more
puzzling why the crossing occurs near the critical point  in Fig. \ref{fidsus_extrap_open}.

Third, we have extrapolated the position of the maximum of fidelity
susceptibility with Eq. (\ref{Jfit}). Using data for $M=16,32,48,64$, we have
obtained from the fit (\ref{Jfit}) that 
$J^*(\infty)/U=0.289\pm0.008$, which again agrees with some
earlier studies of the position of the critical point \cite{Rigol2013}.

Finally, at the risk of stating the obvious, we mention that it is desirable to 
extend these computations to larger (more critical) systems.

\section{Discussion}
\label{sec_discussion}
\begin{figure}
\includegraphics[width=\columnwidth,clip=true]{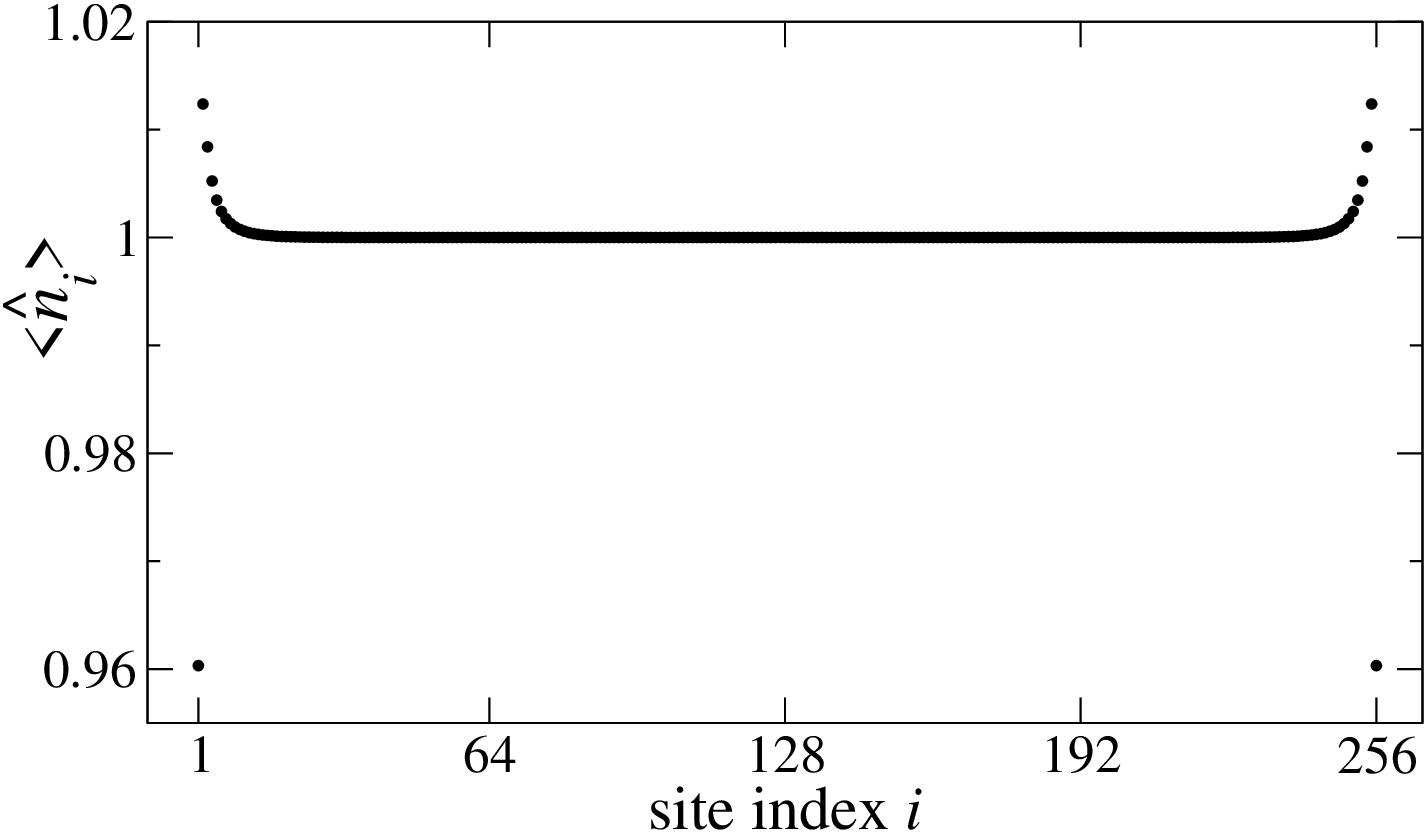}
\caption{Density of atoms $\langle\hat n_i\rangle$ in the ground state of the 
open chain Bose-Hubbard model computed for $M=256$ and $J/U=0.21$, i.e., near the minimum of fidelity 
(Fig. \ref{open_delta0_02_Ndo1024}).
}
\label{gestosc}
\end{figure}

Our results  show that there is a significant difference between
fidelity in the open and periodic chains. The ground states obtained in the two cases 
mainly differ by the occupation of the lattice sites, which is translationally invariant 
in a periodic chain and inhomogeneous near the edges in the open problem (Fig. \ref{gestosc}).
We believe that this inhomogeneity makes the difference, but do not have the
explanation of why it is so large.

It is worth to realize that such an inhomogeneity did not cause much trouble  in the determination of the 
location of the critical point through the studies of the decay of the correlation functions
\cite{MonienLL}. The ground states in these studies were obtained through the open-chain DMRG simulations, 
so they were the same as in our calculations. As expected, the influence of 
inhomogeneities on the system properties near the center  was marginal for large-enough systems.
Thus, one could  obtain reliable results
by computing the two-point correlation functions near the center.

This approach differs from the fidelity approach in  one important aspect.
Namely, the parts of the system near the center and those near the edges equally contribute to  fidelity.
We see  no straightforward way to factor out the influence of the edges on fidelity. 
This complication follows from the simplicity of the fidelity approach, which  democratically
collapses all information about two ground state wave-functions into a
single number.

We would like to stress that better understanding of our numerical
results should come from the analytical derivation of the finite system-size scaling of the position of the 
minimum of fidelity (maximum of fidelity susceptibility) in a BKT transition.
The derivation of the
robust system-size-dependent  scaling expressions  for a BKT transition is notoriously difficult.
This  is seen, e.g.,  from the spread of the estimates of the location
of the critical point. Interestingly enough, our numerics shows 
that  the scaling law capturing the behavior of fidelity 
should distinguish between the open and periodic boundary conditions.
Finally, we mention that our numerically-supported scaling law for fidelity at its
minimum, Eq. (\ref{FJstar}), also awaits analytical  explanation.

Summarizing, the key findings of this manuscript are the following. First, we
have proposed and numerically verified an expression relating 
fidelity at the minimum, the system size, and the parameter shift
(\ref{FJstar}). Second, we have found and numerically characterized the
striking difference between fidelity in the open and periodic chains.
Third, an exact analytical expression for fidelity susceptibility 
in the deep superfluid limit has been derived. 
We expect that these results should 
motivate further studies of fidelity in the Bose-Hubbard model ultimately leading to the
complete understanding of the BKT transition in this model.

\begin{center}
{\bf ACKNOWLEDGMENTS}
\end{center}

We thank Dominique Delande for his comments about the the manuscript and suggesting logarithmic correction formula Eq.~\eqref{Jfit_ln}.
BD and JZ acknowledge support of the Polish National Science Center grant
DEC-2012/04/A/ST2/00088. MŁ acknowledges support 
the Polish National Science Center project no. 2013/08/T/ST2/00112 for the PhD thesis, 
special stipend of Smoluchowski Scientific Consortium ``Matter Energy Future'' and support by Jagiellonian University International Ph.D Studies in Physics of Complex Systems
(Agreement No. MPD/2009/6 with Foundation for Polish
Science). Simulations were 
performed at ACK Cyfronet AGH under the  PL-Grid project and at IF UJ using the  Deszno supercomputer
purchased in the framework of the Polish Innovation Economy Operational Program (POIG.02.01.00-12-023/08). 

\appendix
\section{Fidelity susceptibility at $U/J=0$}
\label{AppA}
We will use Eq.~(\ref{perturb}) to compute fidelity susceptibility at
$\lambda=U/J=0$.
Thus, 
\be
\hat H(U/J=0) = -\sum_{i=1}^M \hat a_i^\dag\hat a_{i+1} + {\rm h.c.}, 
\label{HUJ}
\ee
where $\hat a_{M+1}\equiv\hat a_1$ and 
\bee
\hat V = \frac{1}{2}\sum_{i=1}^M \hat n_i\B{\hat n_i-1}.
\eee
The eigenstates
of the Hamiltonian (\ref{HUJ}) can be obtained in a standard way
\cite{Stoof2001}. By going to the
momentum space 
$$
\hat a_j = \frac{\sum_k \exp(ikj)\hat b_k}{\sqrt{M}}, \
k=0, \frac{2\pi}{M}, \frac{4\pi}{M}, \dots, 2\B{\pi-\frac{\pi}{M}},
$$
one finds that 
\bee
\hat H(U/J=0) = -2\sum_k \cos(k)\hat b^\dag_k\hat b_k.
\eee
All particles occupy the zero 
momentum mode in the  ground state:  $|\phi_0(U/J=0)\rangle=|N,0,\dots\rangle$.
The perturbation operator has to be transformed to the momentum
space as well, and its action on the ground state is the following
\bee
\begin{aligned}
\hat V&|\phi_0(U/J=0)\rangle =\frac{N(N-1)}{2M}|\phi_0(U/J=0)\rangle \\ +
&\frac{\sqrt{N(N-1)}}{2M}
\sum_{k\neq0}\hat b^\dag_{2\pi-k}\hat b^\dag_{k}|N-2,0,\dots\rangle.
\end{aligned}
\eee
One can now use Eq.~(\ref{perturb}) to obtain  
\be
\begin{aligned}
\chi(U/J=0) &= \frac{N(N-1)}{64M^2}\B{\sum_{k'}\frac{1}{\sin^4(k')}+\frac{1}{2}},\\
 k'&=\frac{\pi}{M}, \frac{2\pi}{M},\dots,\frac{\pi}{M}\frac{M-2}{2},
\end{aligned}
\label{s1}
\ee
for even $M$ and 
\be
\begin{aligned}
\chi(U/J=0) &= \frac{N(N-1)}{64M^2}\sum_{k''}\frac{1}{\sin^4(k'')},\\
 k''&=\frac{\pi}{M}, \frac{2\pi}{M},\dots,\frac{\pi}{M}\frac{M-1}{2},
\end{aligned}
\label{s2}
\ee
for odd M.
The sums over $k'$ and $k''$  can be evaluated with the technique proposed in Ref.
\cite{BDfid3}.

The evaluation of the sum (\ref{s1}) proceeds from the identity \cite{Ryzhik}
\be
\begin{aligned}
&\sum_{k'}\BB{\frac{\sin^2(k')}{\sinh(x)}+\frac{\tanh(x/2)}{2} }^{-1} = f(x),\\
&f(x) = M\coth\B{\frac{Mx}{2}} -2\coth(x).
\end{aligned}
\label{ryzhik}
\ee
Dividing this expression by $\sinh(x)$
and then taking its derivative with respect to $x$ we get 
\be
\sum_{k'}\BB{\sin^2(k')+\sinh^2(x/2)}^{-2} =-\frac{2}{\sinh(x)}\frac{d}{dx}\B{\frac{f(x)}{\sinh(x)}}.
\label{qaz}
\ee
The limit of $x\to0$ of Eq.~(\ref{qaz}) provides
\be
\sum_{k'}\frac{1}{\sin^4(k')}= \frac{M^4+10M^2-56}{90}.
\label{wsx}
\ee

One can compute the sum in Eq.~(\ref{s2})  by replacing 
$k'$ by $k''$ and $f(x)$ by  $M\coth(Mx/2) - \coth(x/2)$ in Eq.~(\ref{ryzhik}), and then 
 repeating the steps  leading to Eqs. (\ref{qaz}) and
(\ref{wsx}). In the end,  we get 
\be
\sum_{k''}\frac{1}{\sin^4(k'')}= \frac{M^4+10M^2-11}{90}.
\label{edc}
\ee

Combining Eqs. (\ref{s1}) and  (\ref{wsx}) and also Eqs. (\ref{s2}) and (\ref{edc}),
one obtains 
\bee
\chi(U/J=0) = \frac{N(N-1)}{5760M^2}\B{M^4+10M^2-11},
\eee
for any  system size $M>1$ and the number of atoms $N>1$. 

\section{Convergence of numerical simulations}
\label{AppB}
The accuracy of our numerical computation of fidelity is mainly affected by the accuracy of the
determination of the ground states of the Bose-Hubbard Hamiltonian. 
For the later, we use the DMRG and TEBD algorithms for the open and periodic 
systems, respectively.
Both numerical approaches express the ground state through the
MPS ansatz.

\begin{figure}
\includegraphics[width=\columnwidth,clip=true]{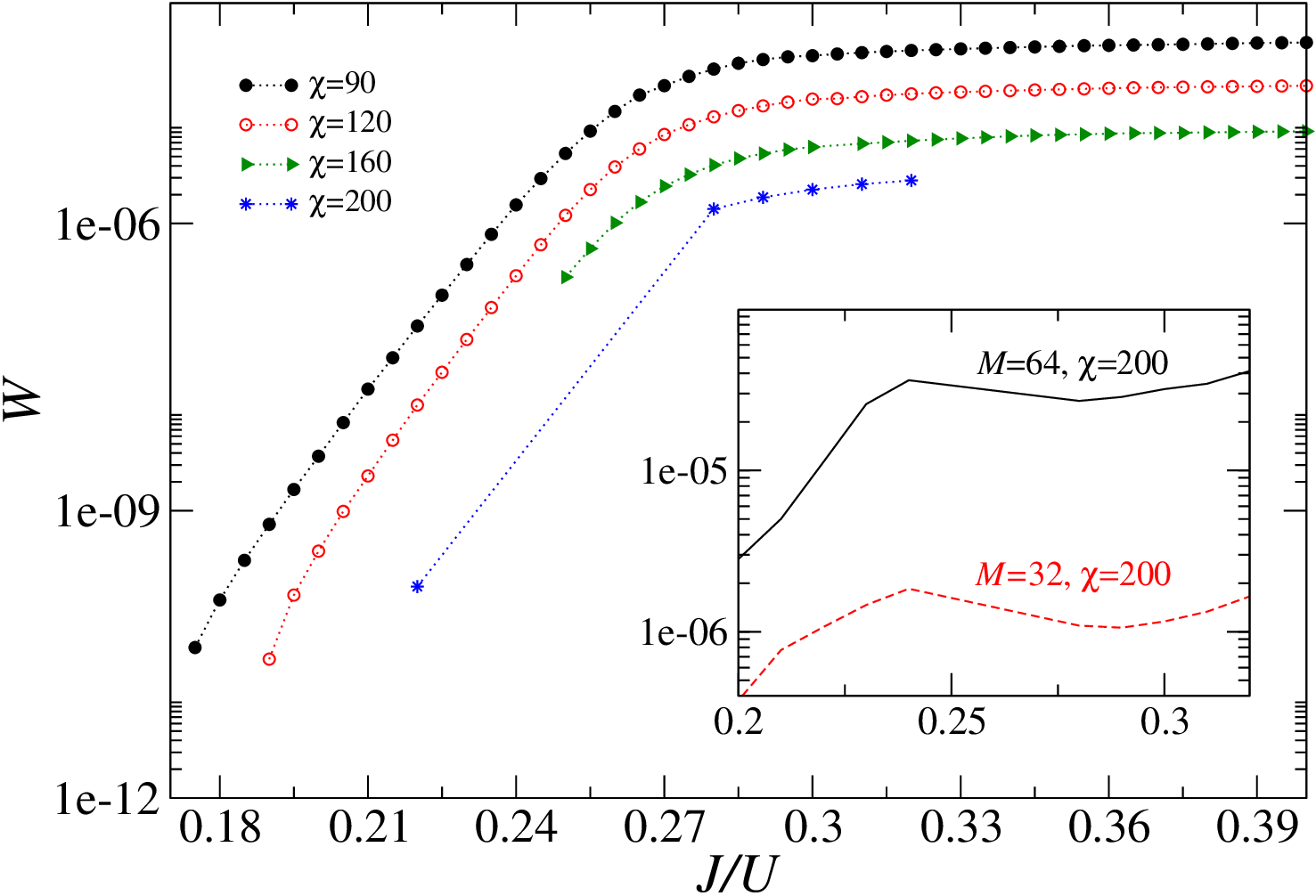}
\caption{(color online) 
Total discarded weights for the computation of ground states. 
The main plot shows data for the open Bose-Hubbard chain of length $M=512$. 
The inset shows the same for the periodic system.
}
\label{dis_w}
\end{figure}

The errors in the DMRG calculations appear mainly  during the so-called local updates, 
where the  predetermined number of the eigenvectors of the reduced density matrix 
with largest eigenvalues $\lambda_i^2$ is kept. 
We keep up to $\chi=200$ eigenvectors out of the superblock 
dimension $d\chi=1400$ ($d$ is the on-site Hilbert space dimension; $\chi$
in this section should not be confused with fidelity susceptibility).  
Let $k$ denote the site at which the local update is considered.
The weight of the discarded  eigenvectors  is  
$$
W_k=\sum\limits_{i=\chi+1}^{d\chi} \lambda_i^2
$$
for that local update. 
Fig.~\ref{dis_w} shows the total discarded weights $W=\sum_k W_k$ as a
function of  the ratio $J/U$ and the bond dimension $\chi$. 

It shows that total discarded weights are certainly negligible around $J/U=0.21$, 
where we study the minima of fidelity (Figs. \ref{Fwmin_0_02_open}--\ref{Jstart_delta0_02_open}).
In particular, we find that for all studied system sizes, $M=32,\dots,2048$,
$F(J/U=0.21,\delta=0.02)$  obtained for $\chi=180$ and $\chi=240$ differ by less than 
$0.001\%$.

On the other hand, for large $J/U$, say $J/U>0.3$
for which the infinite system is superfluid, the discarded
weights are non-negligible. As a result, we find that $1-F(J/U=0.3,\delta=0.02)$  
for  $\chi=160$ and $\chi=200$ differ by about $2\%$ for the system size $M=512$ (we
compare here $1-F$ instead of $F$ because fidelity approaches unity in this
regime of parameters).  This accuracy is sufficient for our qualitative
discussion of fidelity in the superfluid limit.  
Finally, we mention that in the DMRG calculations the sweeps across 
the system have been performed until 
the  energy of the ground state reached a stationary value.

The ground state in the periodic system was obtained by the imaginary time 
evolution using the TEBD algorithm  \cite{Wall2009}. Similarly as in the DMRG
calculation, the maximal  discarded weights can be computed. We plot them 
in the inset of Fig.~\ref{dis_w}. 
Since the eigenvalues $\lambda^2_i$ decrease  much slower than in the periodic case,
we have to focus on small systems. This quick loss of accuracy with the
system size  can be illustrated by computing fidelity for $M=32$ and $M=64$. 
We find that $F(J/U=0.21,\delta=0.02)$ differs for $\chi=180$ and $\chi=240$ 
by about $0.001\%$ when $M=32$. For $M=64$, however, this difference grows to
almost $4\%$. Thus, our results for fidelity in the periodic case 
are accurate near the minimum to a few percent for $M=64$ and to a fraction 
of a percent for smaller system sizes. This is sufficient for our quantitative 
study of the location of the minimum of fidelity.

Finally, we would like to mention that 
the computation of a single ground state in an open (periodic) chain of length
2048 (64) on a single core 2.6 GHz Intel Xeon processor takes us over a month.
Thus, the extension of these studies to larger systems will require the engagement 
of massive computer resources. 

\bibliography{reference}

\begin{thebibliography}{44}
\bibitem[{hom()}]{homo}
\bibinfo{note}{In virtually all experiments the harmonic trapping potential is
  present adding the local density-dependent term to the Hamiltonian. From the
  theoretical point of view, the properties of a homogeneous system are more
  interesting and we shall consider such a system. Recent progress in
  experimental techniques shows that homogeneous models can be experimentally
  realized \cite{Gaunt13}.}
\bibitem[{rem()}]{remark_spline}
\bibinfo{note}{We compute fidelity on the discrete $\Delta J/U=0.005$ grid and
  use the standard cubic spline technique to locate its minimum, i.e.,
  $J^*(M)/U$. Thus, the upper bound on the error bar of $J^*(M)/U$ is $0.005$.
  A more careful estimate of the error bar of each point in Figs.
  \ref{Jstart_delta0_02_open} and \ref{Jstar_delta0_02_periodic} can be
  obtained as follows. First, for every system size that we study, we
  recalculate fidelity on the $\Delta J/U=0.01$ grid, which is twice larger
  then the original one that we employ. Then, we again use the cubic spline
  technique to compute the position of the minimum of fidelity. We find that
  for all system sizes $J^*(M)/U$ computed for $\Delta J/U=0.01$ and $\Delta
  J/U=0.005$ differ by less than $0.0001$. This suggests that the error bars on
  the data points that we use for the fit are negligible.}
\end{thebibliography}
\end{document}